\documentclass[a4paper]{article}

\usepackage{INTERSPEECH2022}

\usepackage{amsmath}
\usepackage{amsthm}
\usepackage{subcaption}
\usepackage{multirow}
\usepackage{amssymb}
\usepackage{pifont}

\usepackage[hidelinks]{hyperref}
\usepackage[usenames,dvipsnames]{pstricks}
\usepackage{pstricks-add}
\usepackage{epsfig}
\usepackage{pst-grad}
\usepackage{pst-plot}
\usepackage{pst-pdf}
\usepackage[space]{grffile}
\usepackage{etoolbox}
\makeatletter
\patchcmd\Gread@eps{\@inputcheck#1 }{\@inputcheck"#1"\relax}{}{}
\makeatother

\newcommand\blfootnote[1]{%
  \begingroup
  \renewcommand\thefootnote{}\footnote{#1}%
  \addtocounter{footnote}{-1}%
  \endgroup
}

\theoremstyle{definition}
\newtheorem{definition}{Definition}[section]

\DeclareMathOperator*{\argmax}{arg\,max}

\title{Extending GCC-PHAT using Shift Equivariant Neural Networks}
\name{Axel Berg$^{1,2}$, Mark O'Connor$^{3, *}$, Kalle Åström$^{2}$, Magnus Oskarsson$^{2}$}

\address{
  $^1$Arm, Sweden, $^2$Lund University, Sweden, $^3$Tenstorrent, Germany}
\email{axel.berg@arm.com, moconnor@tenstorrent.com \\ 
\{karl.astrom, magnus.oskarsson\}@math.lth.se }

\begin{document}

\maketitle
\begin{abstract}
Speaker localization using microphone arrays depends on accurate time delay estimation techniques. For decades, methods based on the generalized cross correlation with phase transform (GCC-PHAT) have been widely adopted for this purpose. Recently, the GCC-PHAT has also been used to provide input features to neural networks in order to remove the effects of noise and reverberation, but at the cost of losing theoretical guarantees in noise-free conditions. We propose a novel approach to extending the GCC-PHAT, where the received signals are filtered using a shift equivariant neural network that preserves the timing information contained in the signals. By extensive experiments we show that our model consistently reduces the error of the GCC-PHAT in adverse environments, with guarantees of exact time delay recovery in ideal conditions. \blfootnote{$^{*}$ Work done while at Arm.}
\end{abstract}

\section{Introduction}

Time delay estimation (TDE) is an essential component in many applications involving acoustic localization, including sound source tracking \cite{diaz2020robust}, robotics \cite{li2016reverberant} and self-calibration \cite{burgess2015toa}. In a typical setup, time delays are estimated by analyzing the signals received by a set of synchronized microphones with known positions. The transmitted waveform, which is assumed to have originated from a single sound source, is considered unknown, as is the time at which it was transmitted. Therefore, the time of travel from the source to the microphones cannot be obtained directly, but instead the time difference of arrival (TDOA) is measured by correlating the received signals. The set of TDOA measurements from a microphone array can then be used to compute the signal direction of arrival (DOA) or the sound source position using multilateration \cite{aastrom2021extension}. 

The generalized cross-correlation (GCC) has been the most widely adopted method for TDOA estimation for many decades. In particular, the phase transform (GCC-PHAT) \cite{knapp1976generalized} filter is commonly used in many acoustic scenarios, due to its fast implementation and robustness in adverse environments. However, with the recent advent of deep learning, a wide variety of methods for sound source localization estimation have been developed without the use of cross-correlations, instead processing only the raw waveforms or spectrograms of the signals \cite{grumiaux2021survey}. Furthermore, several of these methods do not explicitly estimate the TDOAs, but instead train the models to directly predict the DOA \cite{chakrabarty2017broadband, nguyen2021general} or sound source coordinates \cite{vera2018towards}.

Other works have instead attempted to combine GCCs with neural networks \cite{xiao2015learning, he2018deep}. Comanducci et al.\ \cite{comanducci2020time} used an autoencoder network on the outputs of a frequency sliding GCC \cite{cobos2020frequency} in order to de-noise the correlations. Wang et al.\ \cite{wang2021gcc} instead used a neural network for predicting a speech mask that can be interpreted as a learnable frequency-selective linear filter. The speech mask is then applied together with the PHAT when correlating the received signals. Notably, Salvati et al.\ \cite{salvati21_interspeech} proposed computing multiple GCCs, each with its own weighted transfer function, and processing the outputs using a convolutional neural network (CNN) that predicts the TDOA for the two signals. Although this method reduces the average error, it struggles to make accurate prediction within a few samples, which is required for high-precision localization.

We propose a novel method for TDOA estimation by filtering the raw waveforms using a neural network before computing the GCC-PHAT. The network can then be trained to exploit patterns in the data, e.g. the acoustic properties of human speech, in order to remove the effects of noise and reverberation. Furthermore, by using a shift equivariant CNN (SE-CNN), the network can learn to find useful representations while preserving the timing information contained in the signals.

\section{Method}

\begin{figure*}
\resizebox{\linewidth}{!}{
\psscalebox{1.0 1.0}
{
\begin{pspicture}(0,-1.8394854)(17.950586,1.6674266)
\definecolor{colour0}{rgb}{1.0,0.6,0.6}
\definecolor{colour1}{rgb}{0.7019608,0.8,1.0}
\definecolor{colour2}{rgb}{1.0,0.9019608,0.7019608}

\psframe[linecolor=black, linewidth=0.04, fillstyle=solid,fillcolor=colour0, dimen=outer](4.34,1.6674266)(1.56,0.07816184)
\psframe[linecolor=black, linewidth=0.04, fillstyle=solid,fillcolor=colour0, dimen=outer](4.34,-0.2502205)(1.56,-1.8394852)

\rput[bl](2.5835295,0.5628677){\huge $f_{\theta}$}
\rput[bl](2.5835295,-1.3547794){\huge $f_{\theta}$}

\rput[bl](0.011764703,0.7663971){\large $\mathbf{x}_1$}
\rput[bl](0.0,-1.1512499){\large $\mathbf{x}_2$}

\psline[linecolor=black, linewidth=0.04, arrowsize=0.05291667cm 2.0,arrowlength=1.4,arrowinset=0.0]{->}(0.57529414,-1.0018382)(1.4152942,-1.0018382)
\psline[linecolor=black, linewidth=0.04, arrowsize=0.05291667cm 2.0,arrowlength=1.4,arrowinset=0.0]{->}(0.57529414,0.9158089)(1.4152942,0.9158089)

\psframe[linecolor=black, linewidth=0.04, fillstyle=solid,fillcolor=colour1, dimen=outer](9.693529,0.58580893)(6.8235292,-0.9941911)
\rput[bl](7.505884,-0.3241911){GCC-PHAT}

\rput[bl](5.3858824,0.5875736){\large $\mathbf{y}_1$}
\rput[bl](5.3858824,-1.33125){\large $\mathbf{y}_2$}

\psline[linecolor=black, linewidth=0.04, arrowsize=0.05291667cm 2.0,arrowlength=1.4,arrowinset=0.0]{->}(4.4435296,0.9158089)(6.650588,-0.1241911)
\psline[linecolor=black, linewidth=0.04, arrowsize=0.05291667cm 2.0,arrowlength=1.4,arrowinset=0.0]{->}(4.4435296,-1.1794852)(6.630588,-0.3241911)

\psline[linecolor=black, linewidth=0.04, arrowsize=0.05291667cm 2.0,arrowlength=1.4,arrowinset=0.0]{->}(9.850588,-0.1641911)(11.740588,-0.1641911)

\psframe[linecolor=black, linewidth=0.04, fillstyle=solid,fillcolor=colour2, dimen=outer](14.800589,0.5658089)(11.930588,-1.0141912)

\rput[bl](10.5,0.055808906){\large $\mathbf{\tilde{R}}$}

\rput[bl](12.957647,-0.5041911){\huge $g_{\theta'}$}

\psline[linecolor=black, linewidth=0.04, arrowsize=0.05291667cm 2.0,arrowlength=1.4,arrowinset=0.0]{->}(14.950588,-0.1641911)(16.000587,-0.1641911)

\rput[bl](16.110588,-0.3041911){$\mathbf{p}(m | \mathbf{x}_1, \mathbf{x}_2)$}

\end{pspicture}
}
}
\caption{An illustration of our proposed method. The input signals are first filtered using the same neural network $f_{\bm{\theta}}$, then correlated using GCC-PHAT, whose outputs are mapped to a probability distribution over time delays by another neural network $g_{\bm{\theta} '}$.}
\label{overview}
\end{figure*}
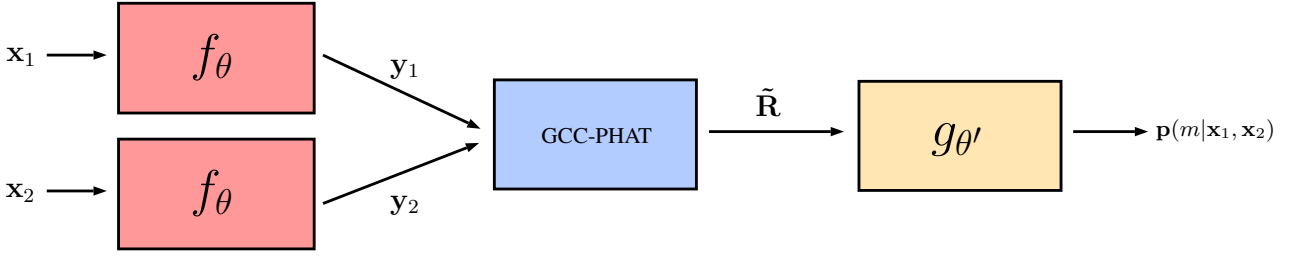

\textbf{Prerequisites.} Consider a reverberant three-dimensional room with two microphones positioned at $\mathbf{r}_1, \mathbf{r}_2 \in \mathbb{R}^3$ and a single sound source positioned at $ \mathbf{r}_s \in \mathbb{R}^3$ emitting an unknown acoustic signal $\mathbf{s}$. Assuming a time-window of $N$ samples, the received signals $\mathbf{x}_1, \mathbf{x}_2 \in \mathbb{R}^N$ at the two microphones can be written as

\begin{gather}
\begin{aligned} 
\label{time}
x_1[n] &= h_1[n] * s[n] + w_1[n], \\
x_2[n] &= h_2[n] * s[n] + w_2[n],
\end{aligned}
\end{gather}
for $n = 0, .., N-1$. Here $h_1[n], h_2[n]$ and $w_1[n], w_2[n]$ are the channel impulse responses from the source to the microphones and additive white noise respectively. Taking the discrete Fourier transform (DFT) of both sides of (\ref{time}) yields

\begin{gather}
\begin{aligned} 
\label{freq}
X_1[k] &= H_1[k] S[k] + W_1[k], \\
X_2[k] &= H_2[k] S[k] + W_2[k],
\end{aligned}
\end{gather}
for $k = 0, ..., N-1$. With this notation, the GCC is defined as

\begin{align}
\label{gcc}
R[m] = \frac{1}{N} \sum_{k=0}^{N-1} \phi[k] X_1[k]X_2^*[k]e^{\frac{i2\pi km}{N}},
\end{align}
for $m = -\tau_{\text{max}}, ..., \tau_{\text{max}}$. The maximal delay is typically taken to be $\tau_{\text{max}} = \lfloor ||\mathbf{r}_1 - \mathbf{r}_2|| F_s / c \rfloor$ where $c$ is the speed of sound and $F_s$ is the sample rate. Furthermore, $\phi[k]$ is a weighting function. In particular, the PHAT \cite{knapp1976generalized} weighting function is given by $\phi[k] = 1/ |X_1[k]X_2^*[k]|$. The estimated time delay is then obtained as

\begin{equation}
\hat{\tau} = \argmax_{m} R[m].
\end{equation}
The PHAT can be regarded as a weighting function that places equal importance on all frequencies in the signal spectrum and only considers the phase of received signals. In an anechoic noise-free environment, the GCC-PHAT outputs a unit impulse centered at the correct time-delay. In the presence of echoes, the PHAT filter attenuates the interference to some extent by limiting the smearing of the correlation, which makes it possible to recover the line of sight component. However, the PHAT also introduces errors from frequency components outside the signal spectrum, which motivates the introduction of a learnable filter that can suppress noise and interference. \\~\

\begin{figure*}
\centering
        \begin{subfigure}[b]{0.3\textwidth}
                \centering
                \includegraphics[width=\linewidth]{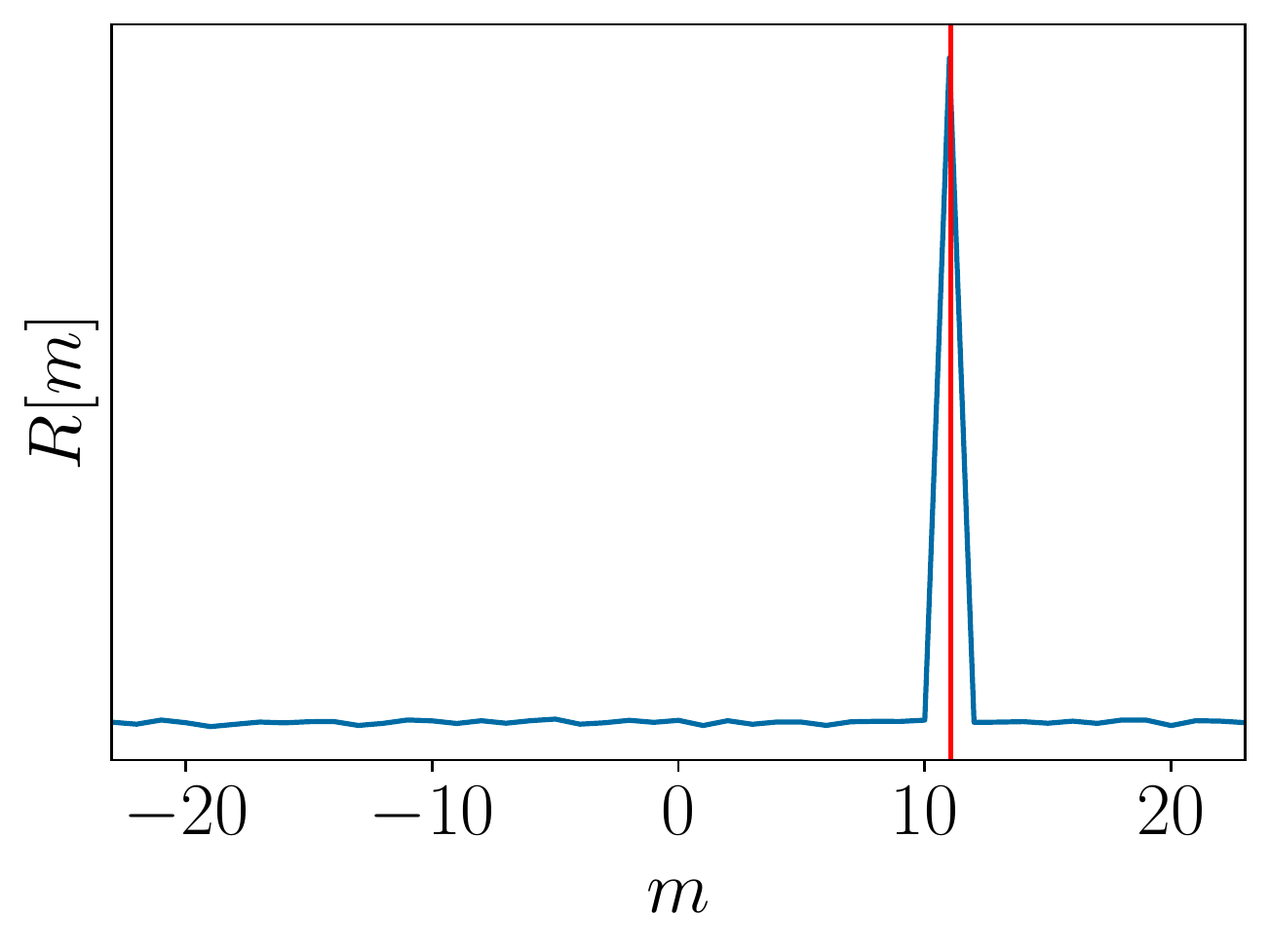}
        \end{subfigure}\hfill
        \begin{subfigure}[b]{0.3\textwidth}
                \centering
                \includegraphics[width=\linewidth]{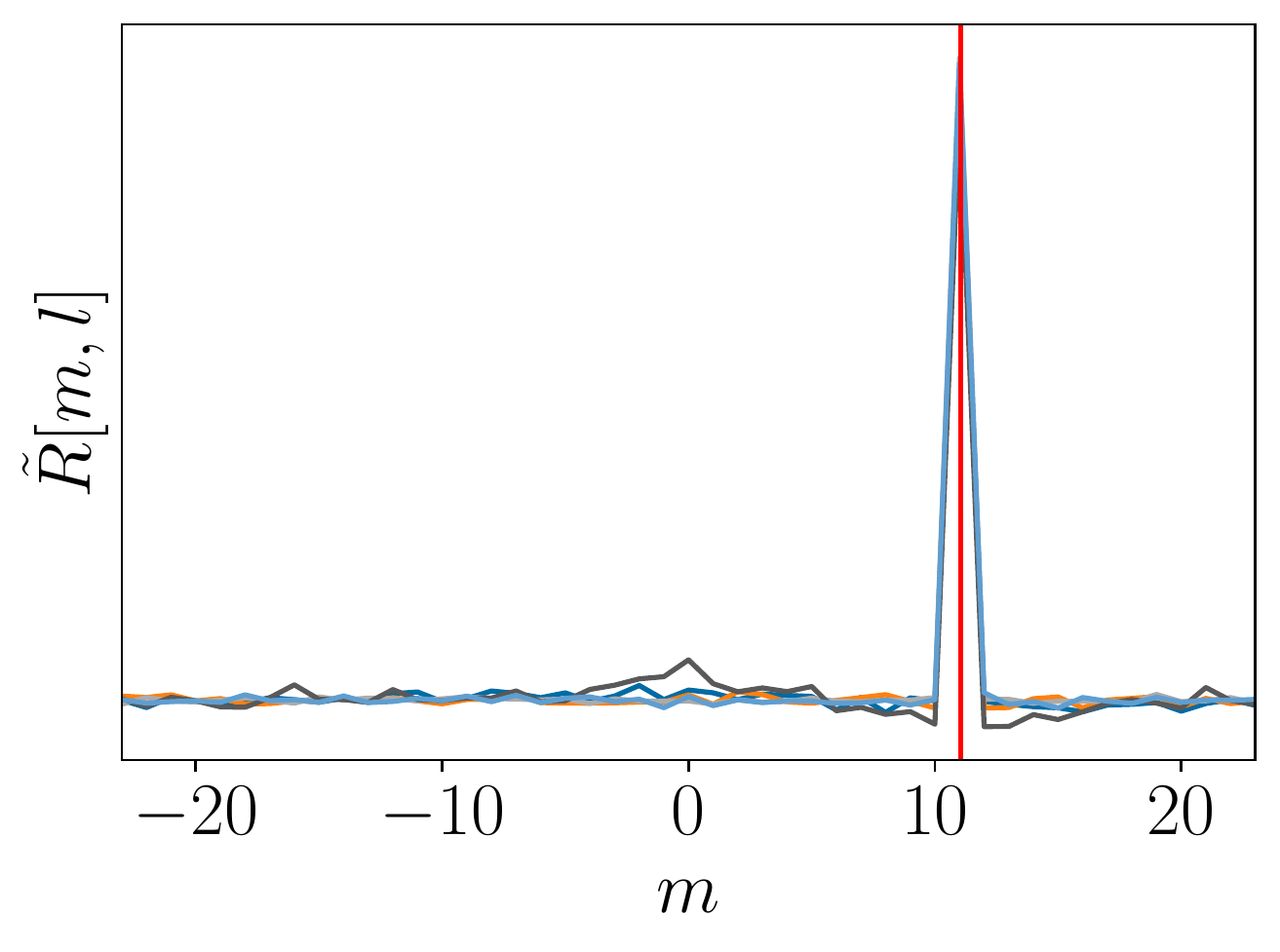}
        \end{subfigure}\hfill
        \begin{subfigure}[b]{0.3\textwidth}
                \centering
                \includegraphics[width=\linewidth]{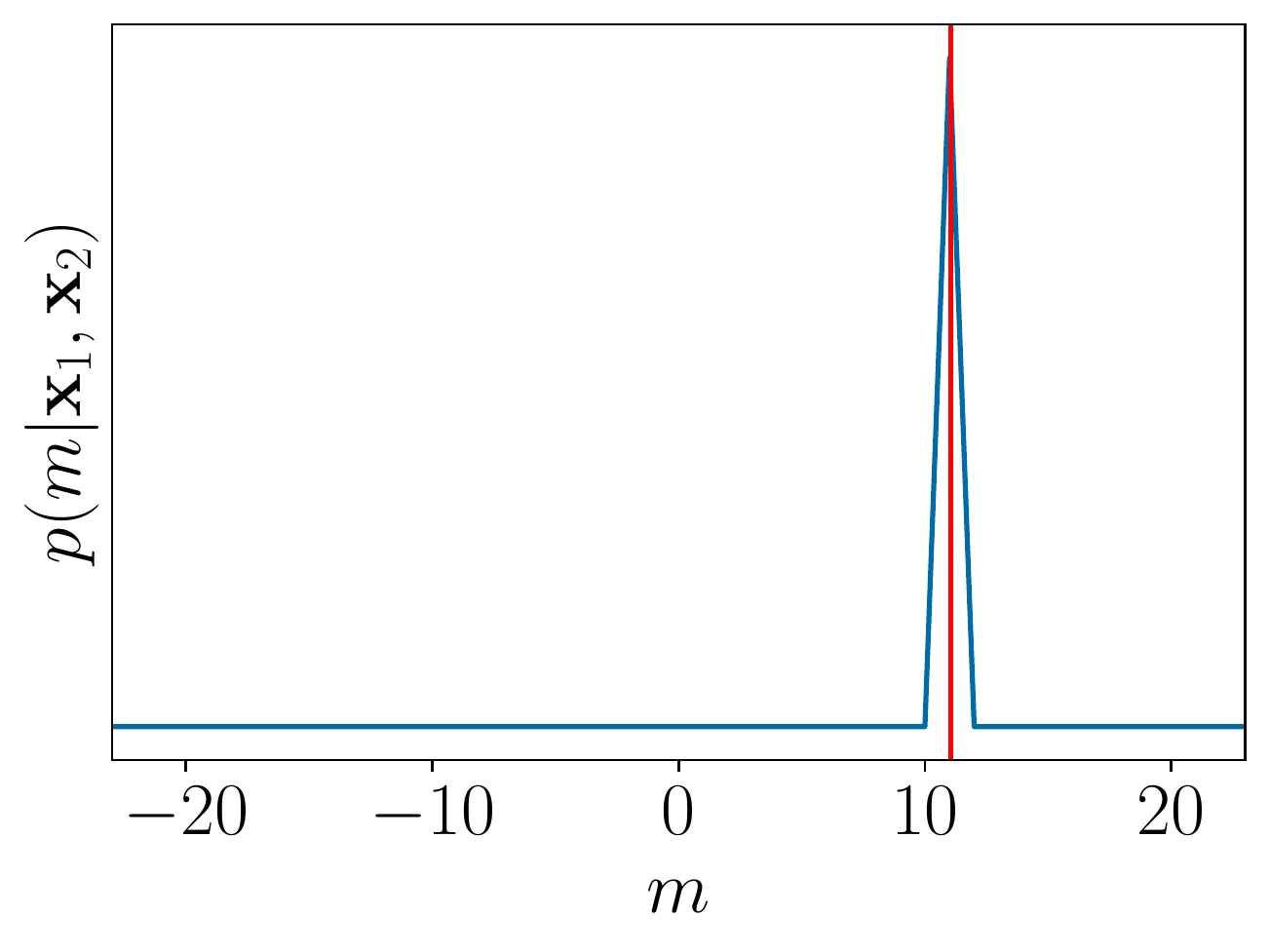}
        \end{subfigure}\hfill
        \begin{subfigure}[b]{0.3\textwidth}
                \centering
                \includegraphics[width=\linewidth]{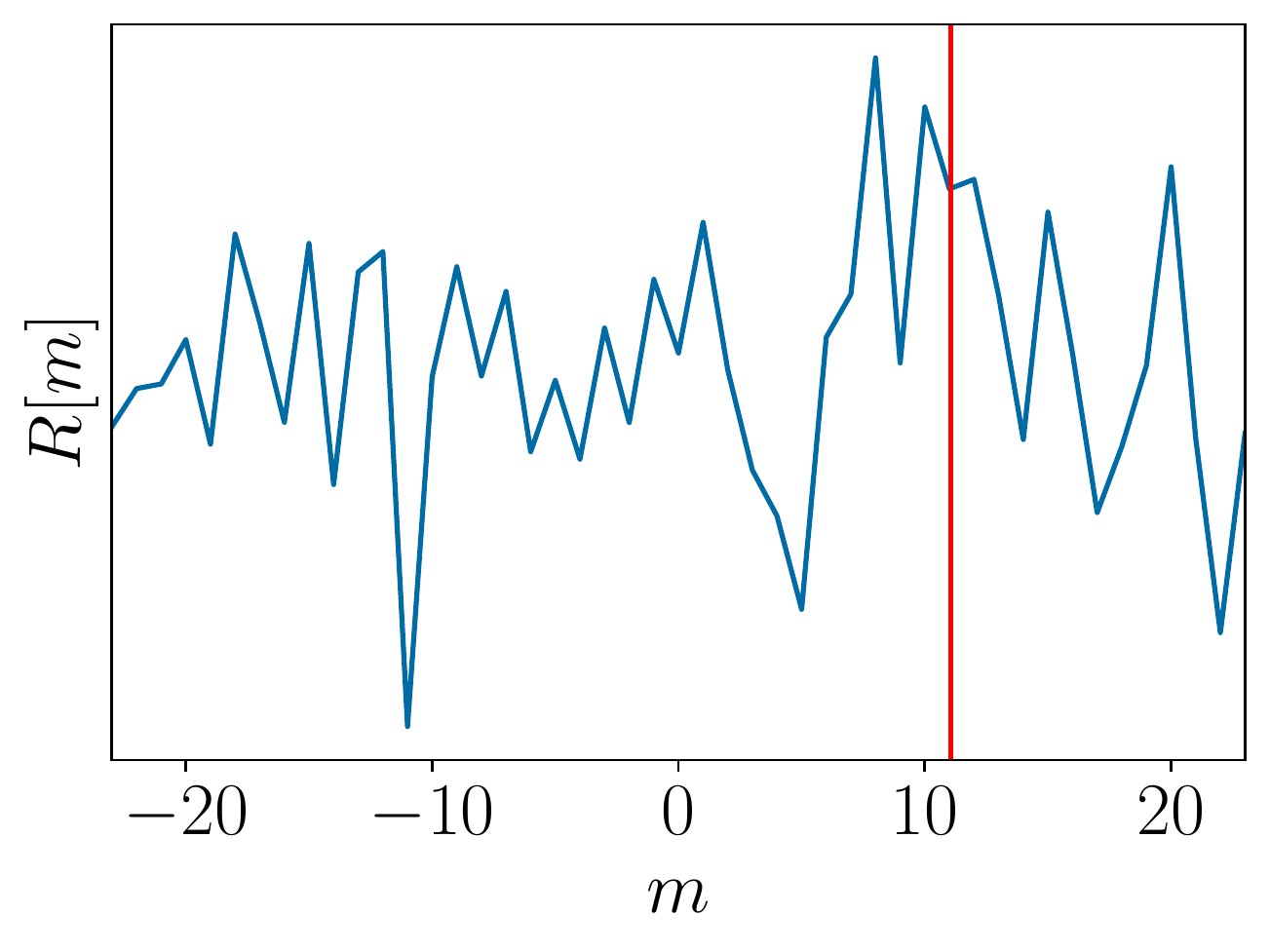}
                \caption{Baseline GCC-PHAT output $\mathbf{R}$.}
        \end{subfigure}\hfill
        \begin{subfigure}[b]{0.3\textwidth}
                \centering
                \includegraphics[width=\linewidth]{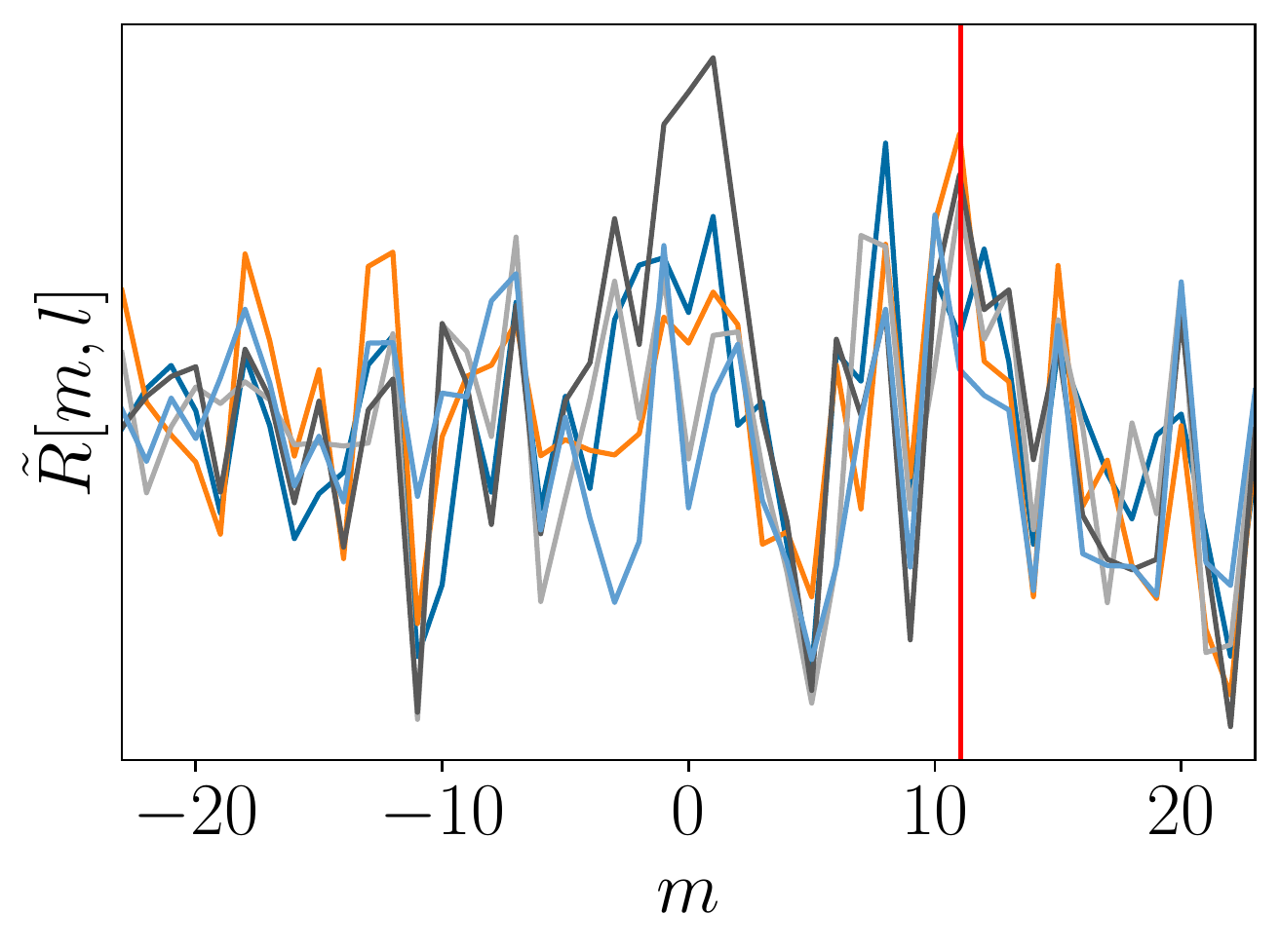}
                \caption{GCC-PHAT output $\tilde{\mathbf{R}}$ of filtered signals.}
        \end{subfigure}\hfill
        \begin{subfigure}[b]{0.3\textwidth}
                \centering
                \includegraphics[width=\linewidth]{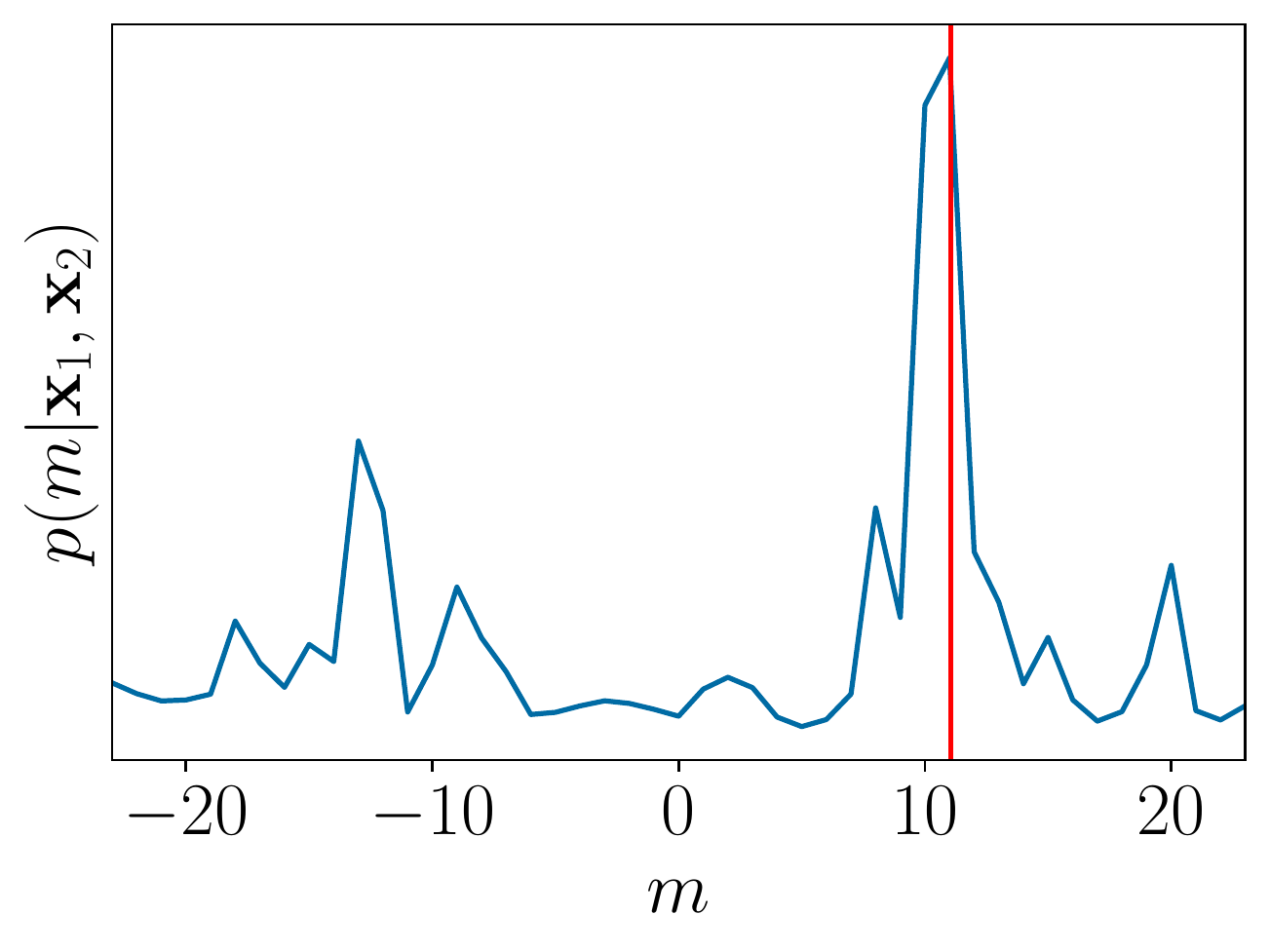}
                \caption{Classifier output $\mathbf{p}(m | \mathbf{x}_1, \mathbf{x}_2)$.}
        \end{subfigure}\hfill
        \caption{Examples of the baseline GCC-PHAT output $\mathbf{R}$ and $\tilde{\mathbf{R}}$ for the unfiltered speech signals $\mathbf{x}_1, \mathbf{x}_2$ and the filtered signals $\mathbf{y}_1, \mathbf{y}_2$ respectively, where the true delay is marked with a red line. For better visualization, only the first five columns of $\tilde{\mathbf{R}}$ are shown. In an ideal environment, the correlations are identical up round-off errors. However, in an adverse environment not all correlations exhibit a clear peak, but the classifier is able to recover the correct delay by combining them into a single probability distribution $\mathbf{p}(m | \mathbf{x}_1, \mathbf{x}_2)$. Top row: ideal noise-free environment. Bottom row: noisy environment.}\label{visualizations}
\end{figure*}

\noindent \textbf{Extending GCC-PHAT.} The main idea of the proposed method is to apply a non-linear filter function on the received signals in order to remove effects of reverberation and noise, which can be realized by using a neural network. However, in the process of doing so, the network needs to preserve the timing information stored in the signals, since this is what we seek to recover when estimating the TDOA. Hence, we seek to employ a network architecture that is explicitly designed for this purpose, and the following definition is then useful.

\begin{definition}
Let $\mathbf{x}$ be a signal and $(\mathbf{x})_\tau$ denote a column-wise circular shift of $\mathbf{x}$. Then $f: \mathbb{R}^N \rightarrow \mathbb{R}^{N \times L}$ is said to be equivariant with respect to time shifts if for any time lag $\tau \in \mathbb{Z}$ we have that $f((\mathbf{x})_\tau) =  (f(\mathbf{x}))_\tau$. 
\end{definition}
\noindent Although CNNs are generally approximately shift-equivariant, not all implementations satisfy this property exactly, due to edge effects. We therefore employ circular padding of the signals before each convolutional layer in the network in order to preserve the timing information completely. For further discussion on convolutions and shift equivariance, we refer to \cite{kayhan2020translation}. 

An overview of the proposed method is illustrated in Figure \ref{overview}. Let $f_{\bm{\theta}}: \mathbb{R}^N \rightarrow \mathbb{R}^{N \times L}$ denote the SE-CNN parameterized by a set of learnable weights $\bm{\theta}$. The network receives input signals of length $N$ and outputs $L$ new signals of the same length 

\begin{gather}
\begin{aligned} 
\mathbf{y}_1 = f_{\bm{\theta}}(\mathbf{x}_1), \\
\mathbf{y}_2 = f_{\bm{\theta}}(\mathbf{x}_2),
\end{aligned}
\end{gather}
where $\mathbf{y}_1, \mathbf{y}_2 \in \mathbb{R}^{N \times L}$, such that each column of the outputs represent a filtered signal. By letting $L > 1$, the network can learn to apply different non-linear filters that capture different properties of the transmitted signals. For each of the $L$ signal components, we then apply the GCC-PHAT individually as

\begin{equation}
\tilde{R}[m, l] = \frac{1}{N} \sum_{k=0}^{N-1} \frac{Y_{1}[k, l]Y_{2}^*[k, l]}{|Y_{1}[k, l]Y_{2}^*[k, l]|}e^{\frac{i2\pi km}{N}},
\end{equation}
where $\mathbf{Y}_1, \mathbf{Y}_2$ are the column-wise DFTs of $\mathbf{y}_1, \mathbf{y}_2$ and $\tilde{\mathbf{R}} \in \mathbb{R}^{(2 \tau_{\text{max}} + 1) \times L}$. Now if $\mathbf{x}_1 = (\mathbf{x}_2)_\tau$, then then it follows from the shift equivariance of the network that  $\mathbf{y}_1 = (\mathbf{y}_2)_\tau$, or equivalently that $Y_1[k, l] = Y_2[k, l] e^{-\frac{i2 \pi k \tau}{N}} $.  Evaluating the GCC-PHAT of the two filtered signals yields

\begin{gather}
\begin{aligned} 
\tilde{R}[m, l] &= \frac{1}{N} \sum_{k=0}^{N-1} \frac{Y_{1}[k, l]Y_{1}^*[k, l]e^{-\frac{i2 \pi k \tau}{N}}}{|Y_{1}[k, l]Y_{1}^*[k, l]e^{-\frac{i2 \pi k \tau}{N}}|}e^{\frac{i2\pi km}{N}} \\
&= \frac{1}{N} \sum_{k=0}^{N-1} e^{\frac{i2 \pi k (m-\tau)}{N}} = \delta_\tau[m],
\end{aligned}
\end{gather}
where $\bm{\delta}_\tau$ is a unit pulse centered at time $\tau$. This shows that in an ideal anechoic noise-free environment, applying $f_{\bm{\theta}}$ to the signals will not prevent the GCC-PHAT from recovering the timing perfectly, as can be seen in Figure \ref{visualizations}b. However, in a reverberant and noisy environment, the network is trained to learn to remove these effects and output signal representations that are equal up to a circular time shift.

In order to output a probability distribution over time shifts, the GCC-PHAT output is fed into another network $g_{\bm{\theta} '}:\mathbb{R}^{(2 \tau_{\text{max}} + 1) \times L} \rightarrow \mathbb{R}^{(2 \tau_{\text{max}} + 1)}$, with its own set of parameters $\bm{\theta}'$ that combines the $L$ different correlations and applies softmax normalization in the final layer. Consequently, the final predictions are obtained as

\begin{equation}
\mathbf{p}(m | \mathbf{x}_1, \mathbf{x}_2) = g_{\bm{\theta} '}(\tilde{\mathbf{R}}),
\end{equation}
such that $\mathbf{p}(m | \mathbf{x}_1, \mathbf{x}_2)$ contains the probabilities for each time delay $m = -\tau_{\text{max}}, ..., \tau_{\text{max}}$ considered in the correlation. As can be seen in Figure \ref{visualizations}c, the trained classifier is able to combine a set of noisy correlations into an accurate prediction. 

The two networks $f_{\bm{\theta}}$ and $g_{\bm{\theta} '}$ can be trained jointly by minimizing the cross-entropy (CE) loss function, which for a single training example becomes

\begin{equation}
L(\mathbf{x}_1, \mathbf{x}_2) = -\sum_{m=-\tau_{\text{max}}}^{\tau_{\text{max}}} \delta_{\tau}[m] \log p(m | \mathbf{x}_1, \mathbf{x}_1),
\end{equation}
where $\tau = \text{round}((||\mathbf{r}_1 - \mathbf{r}_s|| - || \mathbf{r}_2 - \mathbf{r}_s||) F_s / c)$ is the true time delay rounded to the nearest sample. This approach can be regarded as a form of regression-via-classification (RvC), where the model tries to classify the set of correlations into the correct time delay. In contrast, a regression-based approach, such as the one proposed in \cite{salvati21_interspeech}, tries to estimate $\tau$ directly and the model is  trained to minimize the mean squared error (MSE). In Section \ref{experiments} we compare the two methods experimentally in order to justify our RvC approach.

\noindent \textbf{Network architecture.} In order to efficiently process the raw waveforms, we use the SincNet \cite{ravanelli2018speaker} architecture,  which consists of a series of parallel band-pass filters with learnable cutoff frequencies, in the first layer of $f_{\bm{\theta}}$. Specifically, we use 128 filters of length 1023. The following layers consist of regular convolutions with filter lengths of 11, 9 and 7, each with $L = 128$ output channels. Similarly, $g_{\bm{\theta} '}$ consists of 4 convolutional layers with filters of length 11, 9, 7 and 5, all of which has 128 output channels, except for the last layer that has a single output channel which models the log-probabilities for each time delay. In both networks, we use BatchNorm \cite{ioffe2015batch} and LeakyReLU \cite{Maas13rectifiernonlinearities} activations in each layer.

\section{Experiments}
\label{experiments}

\begin{figure*}
\centering
        \begin{subfigure}[t]{0.19\textwidth}
                \centering
                \includegraphics[width=\linewidth, height=\linewidth]{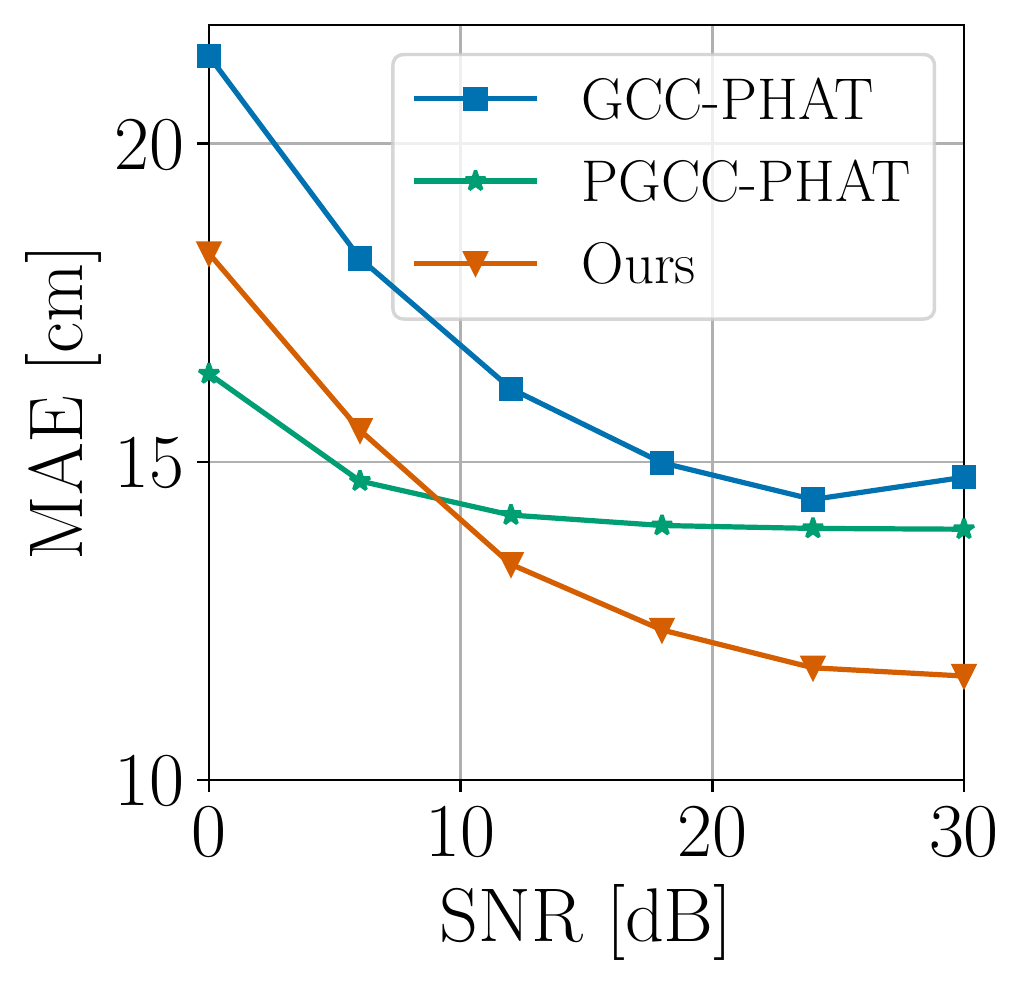}
                \caption{Mean average error for different SNRs}
        \end{subfigure}\hfill
        \begin{subfigure}[t]{0.19\textwidth}
                \centering
                \includegraphics[width=\linewidth, height=\linewidth]{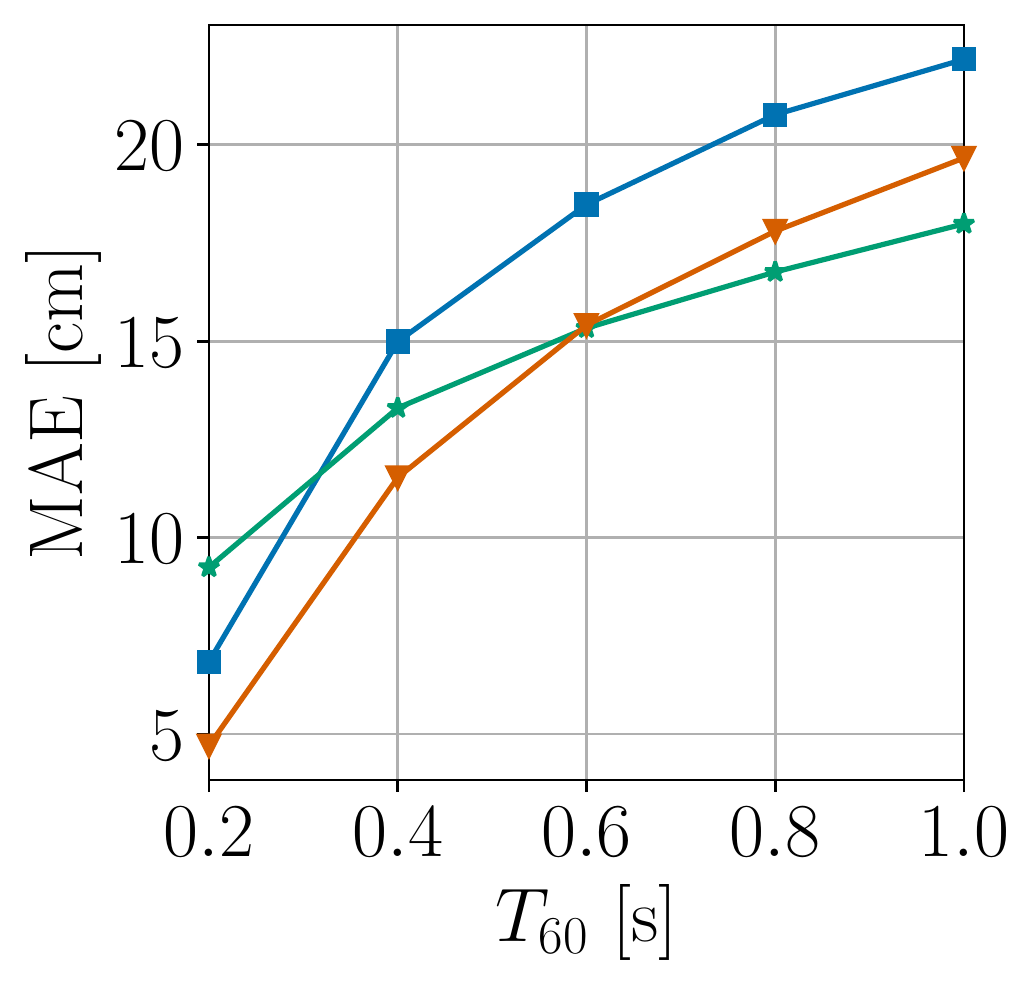}
                \caption{Mean average error for different reverberation times}
        \end{subfigure}\hfill
        \begin{subfigure}[t]{0.19\textwidth}
                \centering
                \includegraphics[width=\linewidth, height=\linewidth]{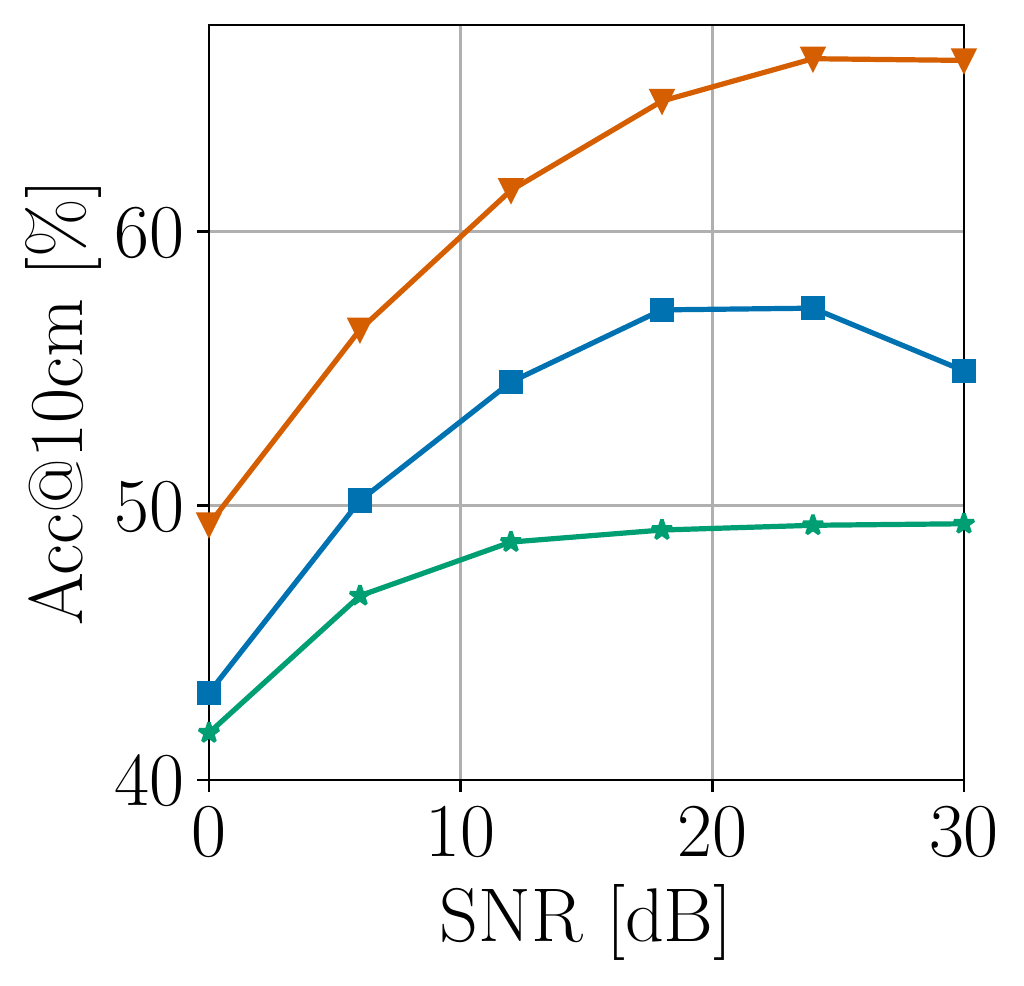}
                \caption{Accuracy for different SNRs}
        \end{subfigure}\hfill
          \begin{subfigure}[t]{0.19\textwidth}
                \centering
                \includegraphics[width=\linewidth, height=\linewidth]{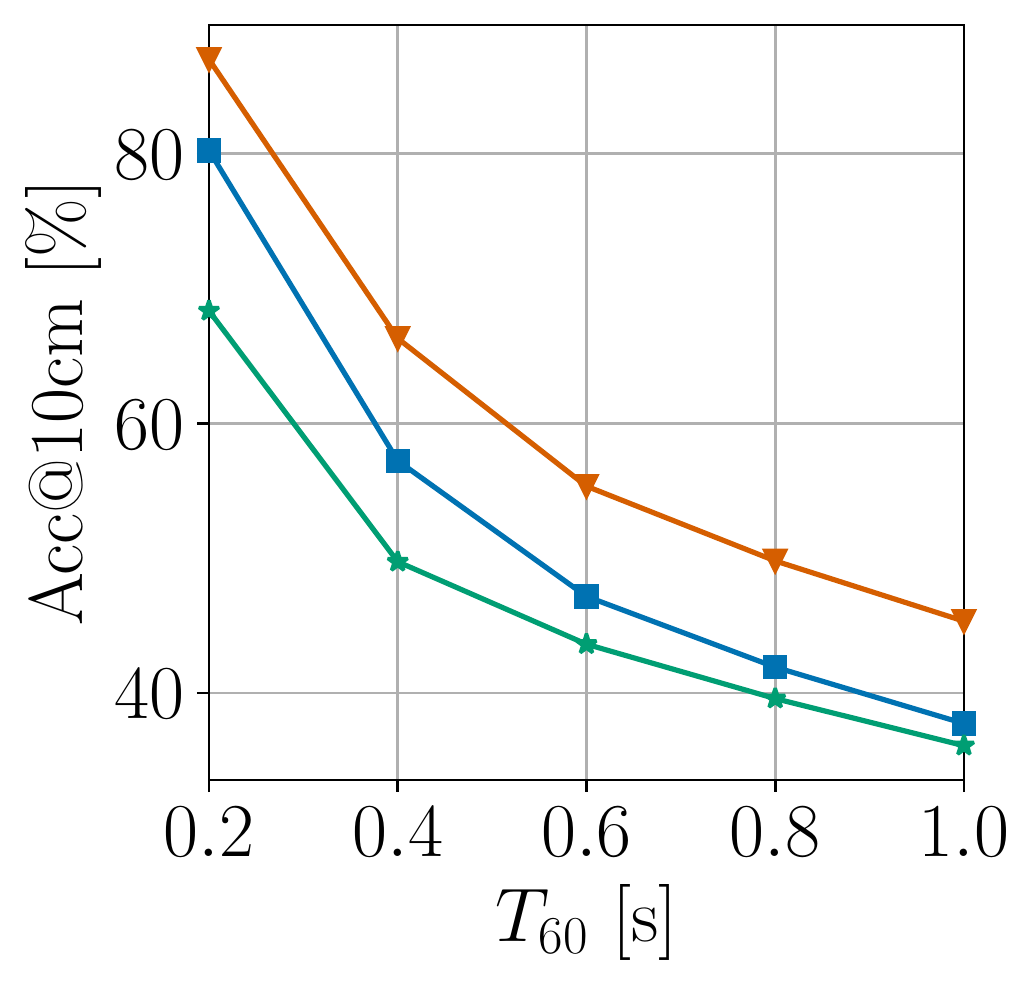}
                \caption{Accuracy for different reverberation times}
        \end{subfigure}\hfill
        \begin{subfigure}[t]{0.19\textwidth}
                \centering
                \includegraphics[width=\linewidth, height=1.0\linewidth]{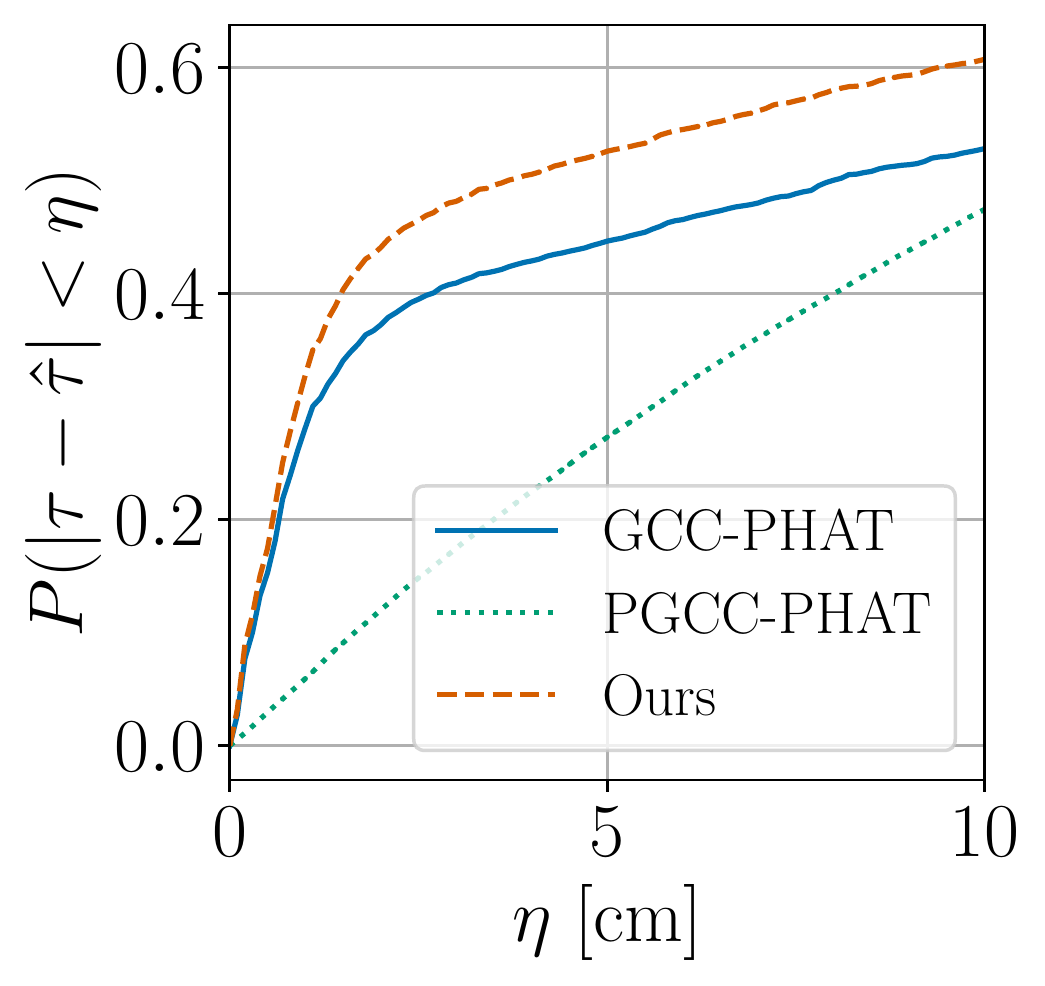}
                \caption{Average probability of a correct detection at various thresholds}
        \end{subfigure}
        \caption{Performance of the different methods in a room of size $6 \times 4 \times 2.5$ m.}
        \label{room1}
\end{figure*}

\begin{figure}[b]
\centering
        \begin{subfigure}[t]{0.33\linewidth}
                \centering
                \includegraphics[width=\linewidth]{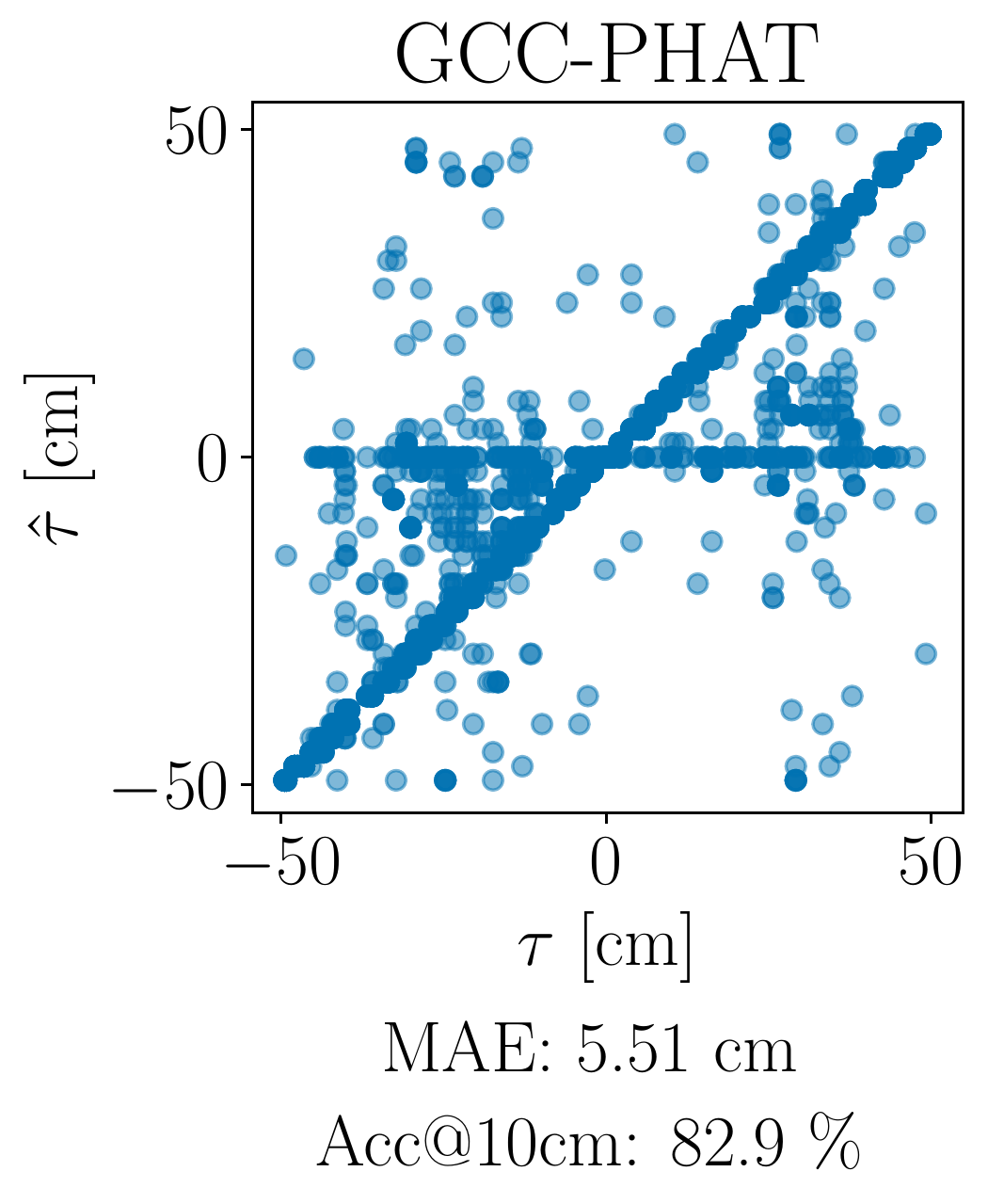}
        \end{subfigure}\hfill
        \begin{subfigure}[t]{0.33\linewidth}
                \centering
                \includegraphics[width=\linewidth]{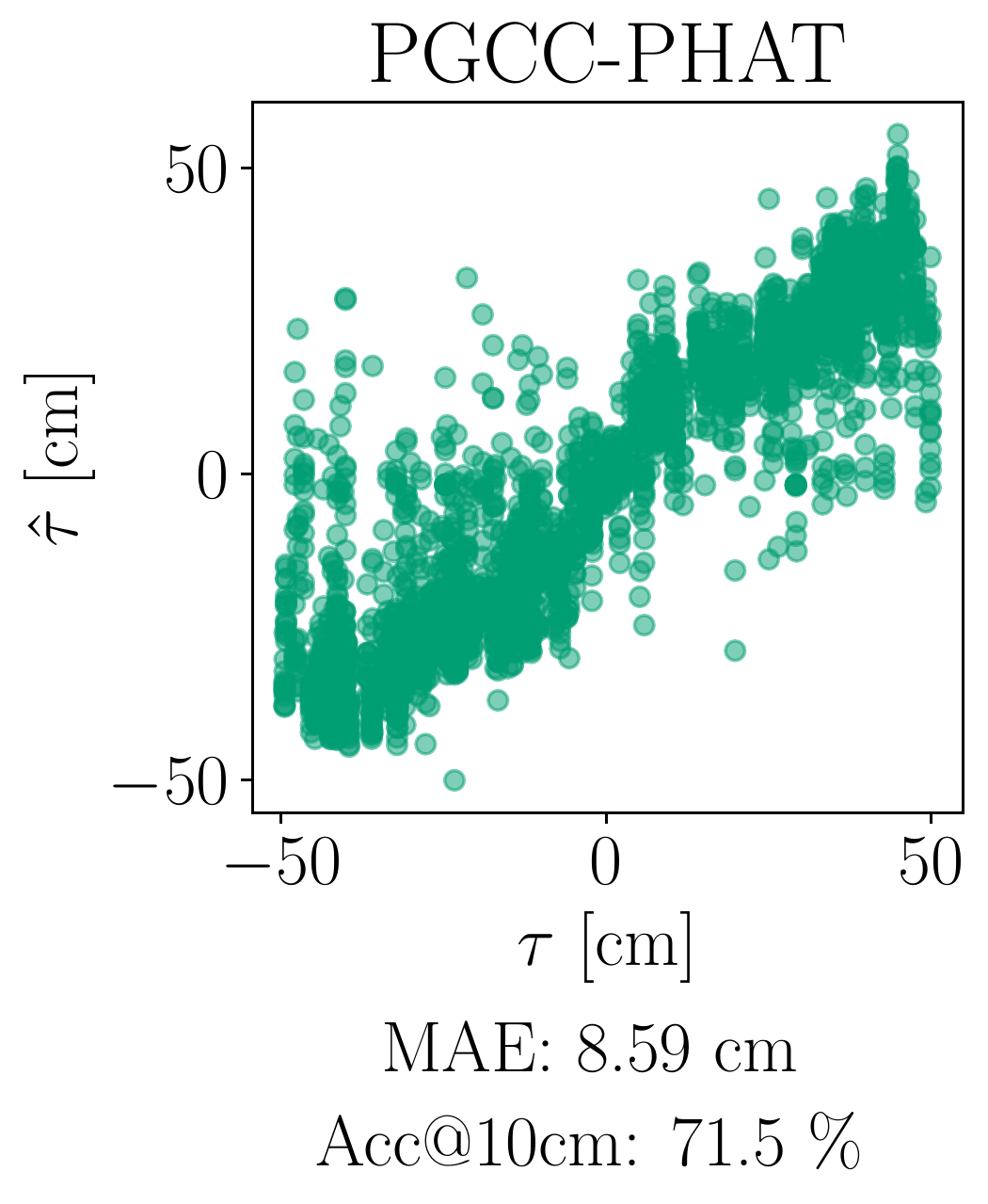}
        \end{subfigure}\hfill
        \begin{subfigure}[t]{0.33\linewidth}
                \centering
                \includegraphics[width=\linewidth]{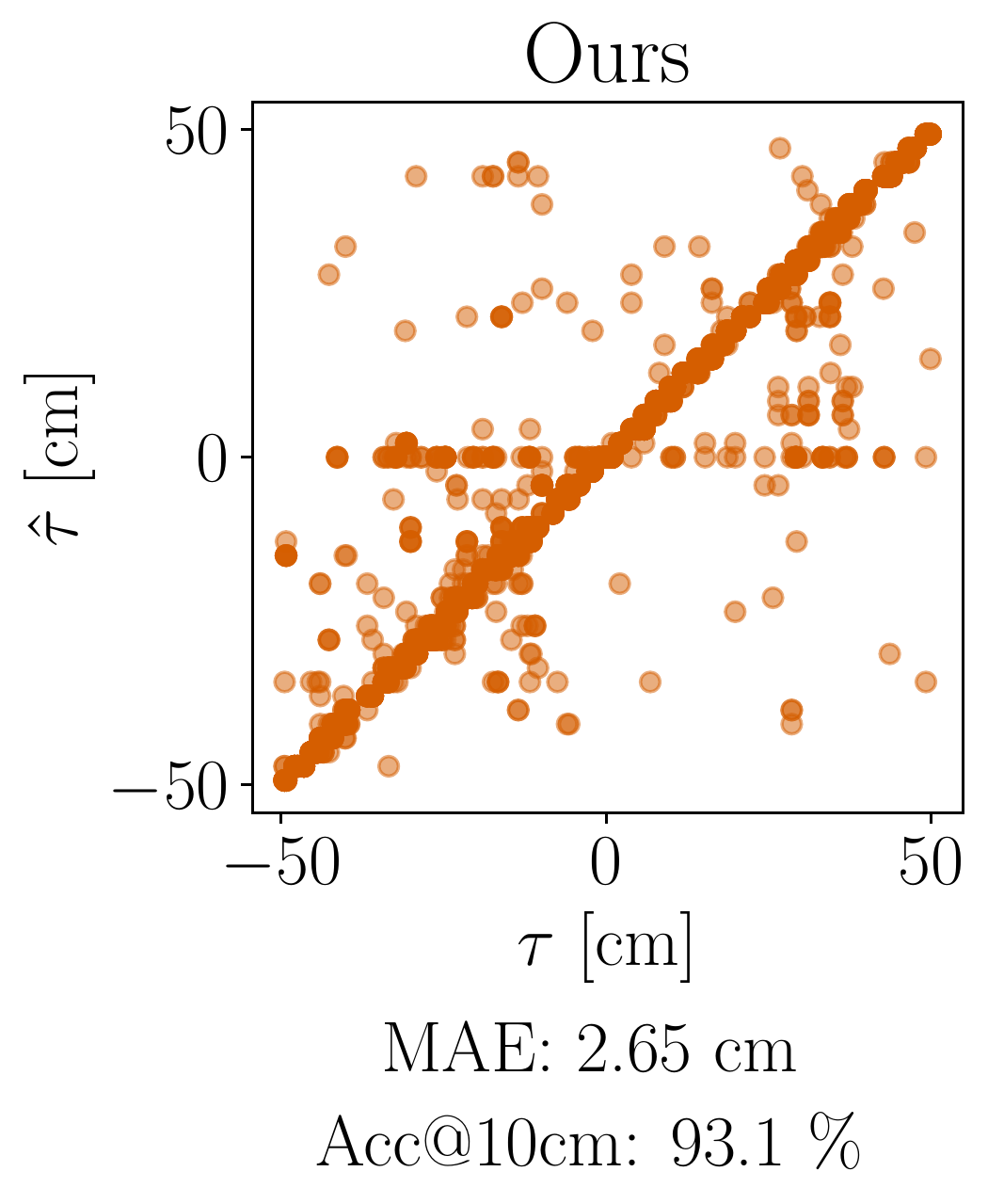}
        \end{subfigure}\hfill
        \caption{Scatter plots of ground truth and predicted time delays for reverberation time $T_{60} = 0.2$ s and SNR = 30 dB.}
        \label{scatter}
\end{figure}

We perform a series of simulated experiments in order to evaluate our method and compare it to baselines. In order to simulate realistic sound propagation, we use Pyroomacoustics \cite{scheibler2018pyroomacoustics}, which enables modeling of reverberant indoor environments based on the image source method \cite{allen1979image}. The audio signals were collected from the LibriSpeech dataset \cite{panayotov2015librispeech}, which contains speech recordings from read audiobooks in English, sampled at $F_s = 16$ kHz. We split the data based on speakers, such that 40 speakers were used for training, 3 for validation and 3 for testing. For each recording we first remove silent parts using a voice activity detector and then extract a 2 second long snippet from each recording. This results in 1892 snippets for training (corresponding to roughly one hour of audio), 188 for validation and 216 for testing. During training, we randomly sample a frame of $N = 2048$ samples for each snippet, while during testing we evaluate on each of 15 non-overlapping windows, for a total of $216 \cdot 15 = 3240$ time delay estimates.

Network training was done inside a simulated room of dimension $7 \times 5 \times 3$ m, with microphones placed roughly in the middle of the room at $\mathbf{r}_1 = [3.5, 2.25, 1.5]^T$ and ${r}_2 = [3.5, 2.75, 1.5]^T$ m from the origin. This setup results in a maximum delay of $\tau_{\text{max}} = 23$ samples. The source positions $\mathbf{r}_s$ were sampled from randomly from a uniform distribution over the entire room for each training sample. Furthermore, random reverberation times $T_{60}$ and signal-to-noise ratios (SNR) were uniformly sampled in the ranges [0.2, 1.0] s and [0, 30 dB] respectively. We use the Adam optimizer \cite{adam} with a batch size of 32, a learning rate of 0.001 with a cosine decay schedule and train the network for 30 epochs.\footnote{Code available at: \url{https://github.com/axeber01/ngcc}}

For comparison, we implement the PGCC-PHAT method following the description in \cite{salvati21_interspeech}. The method uses a CNN that takes several differently weighted GCC-PHAT correlations as input and combines them into a single time delay prediction. Each correlation has a different filter $\phi_\beta[k] = 1/ |X_1[k]X_2^*[k]|^\beta$, for $\beta \in \{0, 0.1, ..., 1\}$. In contrast to our method, the network is trained to minimize the MSE loss.

The trained models were evaluated in a different room with dimensions $6 \times 4 \times 2.5$ m and with the microphones placed at $\mathbf{r}_1 = [3, 1.75, 1.25]^T$ and $\mathbf{r}_2 = [3, 2.25, 1.25]^T$ m respectively, and the source positions were again sampled randomly across the whole room. Each recording was evaluated for SNRs $\in \{0, 6, 12, 18, 24, 30\}$ dB and reverberation times $T_{60} \in \{0.2, 0.4, 0.6, 0.8, 1\}$ s. 

The results are presented in Figure \ref{room1}, where the mean average error (MAE) and accuracy is presented for different SNRs and reverberation times. All TDOA errors have been converted to their corresponding distance errors and the accuracy should be interpreted as the probability $P(|\tau - \hat{\tau}| < \eta)$, where $\eta = 10$ cm, since this a typical level of average precision that can be achieved by an acoustic localization system \cite{omologo1997use, wang2007robust}. For completeness, we show results for lower thresholds in Figure \ref{room1}e as well. Our method achieves the highest accuracy in all conditions and the lowest MAE for conditions with high SNR or low reverberation time, consistently outperforming the GCC-PHAT baseline. A comparison of error distributions for the different methods in a high SNR environment can be seen in Figure \ref{scatter}. Since PGCC-PHAT has been trained to minimize the MSE, its error distribution has a smaller tail but fails to make accurate predictions within a few centimeters, which is necessary for real-world indoor localization. The large number of predictions at $\hat{\tau} = 0$ is due to imperfect audio pre-processing, which results in some time windows containing long periods of silence.

In order to disentangle the influence of the CE and MSE loss functions, we ablate the two by changing only the last layer of the networks. In Table \ref{ablation} it can be seen that the higher accuracy of our method cannot be attributed solely to the CE loss, since it is more accurate than PGCC-PHAT when trained with the same loss function. However, our method is not as effective when trained with the MSE loss. We attribute this to the fact that in contrast to PGCC-PHAT, the entries of our correlation matrix $\tilde{\mathbf{R}}$ do not exhibit any time-smearing, which causes the correlation peaks to be uniformly distributed for incorrect detections. Nevertheless, the RvC approach is preferable for both methods when accuracy is most important. Additionally, the smaller number of parameters in our model makes it more feasible for use in efficient real-time inference on small devices.

In order to demonstrate the effectiveness of using multiple GCC-PHAT correlations, we ablate the number of channels $L$ in the last layer of our feature extractor $f_{\bm{\theta}}$, leaving the other layers unchanged. In Table \ref{ablation2}, it can be seen that adding more channels yields better performance. Moreover, replacing the SincNet layer with a standard learnable convolution with the same filter length results in a performance drop, which motivates the use of learnable bandpass filters in the first layer.

\begin{table}
\footnotesize
  \caption{Results using different loss functions for reverberation time $T_{60} = 0.2$ s, averaged over all SNRs.}
  \label{ablation}
  \centering
  \begin{tabular}{l|c|cc|cc}
    \toprule
    \textbf{Model}  & \textbf{GCC-PHAT} & \multicolumn{2}{c|}{\textbf{PGCC-PHAT}} & \multicolumn{2}{c}{\textbf{Ours}} \\
    \midrule
    \textbf{\#params}  & 0 & \multicolumn{2}{c}{11.5M} & \multicolumn{2}{c}{0.9M} \\
    \textbf{Loss} & - & MSE & CE & MSE & CE \\
    \midrule
    RMSE [cm] & 15.21 & \textbf{10.87} & 13.61 & 12.44 & 13.24  \\
    MAE [cm] & 6.84 & 6.59 & 5.39 & 8.38 & \textbf{5.13}  \\
    Acc@10cm & 80.2 & 80.6 & 85.7 & 71.9 & \textbf{86.5} \\
	\bottomrule
  \end{tabular}
\end{table}

\begin{table}
\footnotesize
  \caption{Results for our method using different number of correlation channels $L$ for reverberation time $T_{60} = 0.2$ s, averaged over all SNRs.}
  \label{ablation2}
  \centering
  \begin{tabular}{l|cccc|c}
    \toprule
    \textbf{\#channels} $L$  & $\mathbf{1}$ & $\mathbf{8}$ & $\mathbf{32}$ & $\mathbf{128}$ & $\mathbf{128}$ w/o SincNet \\
    \midrule
    MAE [cm] & 5.59 & 5.35 & 5.23 & \textbf{5.13} & 5.38 \\
    Acc@10cm & 84.4 & 85.6  & 85.8 & \textbf{86.5} & 85.4 \\
	\midrule
  \end{tabular}
\end{table}

\section{Conclusions}

We have demonstrated that our proposed method is able to consistently improve detection accuracy over the baseline GCC-PHAT and PGCC-PHAT. Furthermore, as the signal strength increases relative to noise and echoes, in the limit our method is guaranteed to recover the time delay within sample accuracy. Although incorrect detections sometimes results in large errors, this is of less practical importance, since robust localization methods are designed to handle a large fraction of incorrect detections by removing outliers in the measurements \cite{aastrom2021extension, plinge2016acoustic, velasco2016tdoa}. Moreover, it can be used as a drop-in replacement for GCC-PHAT, and the outputs from the SE-CNN can be re-used when considering more than two microphones. We therefore conclude that our method would be a suitable alternative for time delay estimation in real-world speaker localization scenarios. 

In future work, we will consider integrating our method into a full sound source localization system. This requires tracking delays over time, as well as considering the geometry of the microphones and sound sources. Here we see further potential of applying machine learning methods in order to tackle the localization problem with an end-to-end approach.

\section{Acknowledgments}

This work was partially supported by ELIIT and the Wallenberg AI, Autonomous Systems and Software Program (WASP), funded by the Knut and Alice Wallenberg Foundation.

\clearpage
\newpage
\bibliographystyle{IEEEtran}

\bibliography{tdoa}

\begin{thebibliography}{10}
\providecommand{\url}[1]{#1}
\csname url@samestyle\endcsname
\providecommand{\newblock}{\relax}
\providecommand{\bibinfo}[2]{#2}
\providecommand{\BIBentrySTDinterwordspacing}{\spaceskip=0pt\relax}
\providecommand{\BIBentryALTinterwordstretchfactor}{4}
\providecommand{\BIBentryALTinterwordspacing}{\spaceskip=\fontdimen2\font plus
\BIBentryALTinterwordstretchfactor\fontdimen3\font minus
  \fontdimen4\font\relax}
\providecommand{\BIBforeignlanguage}[2]{{%
\expandafter\ifx\csname l@#1\endcsname\relax
\typeout{** WARNING: IEEEtran.bst: No hyphenation pattern has been}%
\typeout{** loaded for the language `#1'. Using the pattern for}%
\typeout{** the default language instead.}%
\else
\language=\csname l@#1\endcsname
\fi
#2}}
\providecommand{\BIBdecl}{\relax}
\BIBdecl

\bibitem{diaz2020robust}
D.~Diaz-Guerra, A.~Miguel, and J.~R. Beltran, ``Robust sound source tracking
  using srp-phat and 3d convolutional neural networks,'' \emph{IEEE/ACM
  Transactions on Audio, Speech, and Language Processing}, vol.~29, pp.
  300--311, 2020.

\bibitem{li2016reverberant}
X.~Li, L.~Girin, F.~Badeig, and R.~Horaud, ``Reverberant sound localization
  with a robot head based on direct-path relative transfer function,'' in
  \emph{2016 IEEE/RSJ International Conference on Intelligent Robots and
  Systems (IROS)}.\hskip 1em plus 0.5em minus 0.4em\relax IEEE, 2016, pp.
  2819--2826.

\bibitem{burgess2015toa}
S.~Burgess, Y.~Kuang, and K.~{\AA}str{\"o}m, ``Toa sensor network
  self-calibration for receiver and transmitter spaces with difference in
  dimension,'' \emph{Signal Processing}, vol. 107, pp. 33--42, 2015.

\bibitem{aastrom2021extension}
K.~{\AA}str{\"o}m, M.~Larsson, G.~Flood, and M.~Oskarsson, ``Extension of
  time-difference-of-arrival self calibration solutions using robust
  multilateration,'' in \emph{2021 29th European Signal Processing Conference
  (EUSIPCO)}.\hskip 1em plus 0.5em minus 0.4em\relax IEEE, 2021, pp. 870--874.

\bibitem{knapp1976generalized}
C.~Knapp and G.~Carter, ``The generalized correlation method for estimation of
  time delay,'' \emph{IEEE transactions on acoustics, speech, and signal
  processing}, vol.~24, no.~4, pp. 320--327, 1976.

\bibitem{grumiaux2021survey}
P.-A. Grumiaux, S.~Kiti{\'c}, L.~Girin, and A.~Gu{\'e}rin, ``A survey of sound
  source localization with deep learning methods,'' \emph{arXiv preprint
  arXiv:2109.03465}, 2021.

\bibitem{chakrabarty2017broadband}
S.~Chakrabarty and E.~A. Habets, ``Broadband doa estimation using convolutional
  neural networks trained with noise signals,'' in \emph{2017 IEEE Workshop on
  Applications of Signal Processing to Audio and Acoustics (WASPAA)}.\hskip 1em
  plus 0.5em minus 0.4em\relax IEEE, 2017, pp. 136--140.

\bibitem{nguyen2021general}
T.~N.~T. Nguyen, N.~K. Nguyen, H.~Phan, L.~Pham, K.~Ooi, D.~L. Jones, and W.-S.
  Gan, ``A general network architecture for sound event localization and
  detection using transfer learning and recurrent neural network,'' in
  \emph{ICASSP 2021-2021 IEEE International Conference on Acoustics, Speech and
  Signal Processing (ICASSP)}.\hskip 1em plus 0.5em minus 0.4em\relax IEEE,
  2021, pp. 935--939.

\bibitem{vera2018towards}
J.~M. Vera-Diaz, D.~Pizarro, and J.~Macias-Guarasa, ``Towards end-to-end
  acoustic localization using deep learning: From audio signals to source
  position coordinates,'' \emph{Sensors}, vol.~18, no.~10, p. 3418, 2018.

\bibitem{xiao2015learning}
X.~Xiao, S.~Zhao, X.~Zhong, D.~L. Jones, E.~S. Chng, and H.~Li, ``A
  learning-based approach to direction of arrival estimation in noisy and
  reverberant environments,'' in \emph{2015 IEEE International Conference on
  Acoustics, Speech and Signal Processing (ICASSP)}.\hskip 1em plus 0.5em minus
  0.4em\relax IEEE, 2015, pp. 2814--2818.

\bibitem{he2018deep}
W.~He, P.~Motlicek, and J.-M. Odobez, ``Deep neural networks for multiple
  speaker detection and localization,'' in \emph{2018 IEEE International
  Conference on Robotics and Automation (ICRA)}.\hskip 1em plus 0.5em minus
  0.4em\relax IEEE, 2018, pp. 74--79.

\bibitem{comanducci2020time}
L.~Comanducci, M.~Cobos, F.~Antonacci, and A.~Sarti, ``Time difference of
  arrival estimation from frequency-sliding generalized cross-correlations
  using convolutional neural networks,'' in \emph{ICASSP 2020-2020 IEEE
  International Conference on Acoustics, Speech and Signal Processing
  (ICASSP)}.\hskip 1em plus 0.5em minus 0.4em\relax IEEE, 2020, pp. 4945--4949.

\bibitem{cobos2020frequency}
M.~Cobos, F.~Antonacci, L.~Comanducci, and A.~Sarti, ``Frequency-sliding
  generalized cross-correlation: A sub-band time delay estimation approach,''
  \emph{IEEE/ACM Transactions on Audio, Speech, and Language Processing},
  vol.~28, pp. 1270--1281, 2020.

\bibitem{wang2021gcc}
J.~Wang, X.~Qian, Z.~Pan, M.~Zhang, and H.~Li, ``Gcc-phat with speech-oriented
  attention for robotic sound source localization,'' in \emph{2021 IEEE
  International Conference on Robotics and Automation (ICRA)}.\hskip 1em plus
  0.5em minus 0.4em\relax IEEE, 2021, pp. 5876--5883.

\bibitem{salvati21_interspeech}
D.~Salvati, C.~Drioli, and G.~L. Foresti, ``{Time Delay Estimation for Speaker
  Localization Using CNN-Based Parametrized GCC-PHAT Features},'' in
  \emph{Proc. Interspeech 2021}, 2021, pp. 1479--1483.

\bibitem{kayhan2020translation}
O.~S. Kayhan and J.~C.~v. Gemert, ``On translation invariance in cnns:
  Convolutional layers can exploit absolute spatial location,'' in
  \emph{Proceedings of the IEEE/CVF Conference on Computer Vision and Pattern
  Recognition}, 2020, pp. 14\,274--14\,285.

\bibitem{ravanelli2018speaker}
M.~Ravanelli and Y.~Bengio, ``Speaker recognition from raw waveform with
  sincnet,'' in \emph{2018 IEEE Spoken Language Technology Workshop
  (SLT)}.\hskip 1em plus 0.5em minus 0.4em\relax IEEE, 2018, pp. 1021--1028.

\bibitem{ioffe2015batch}
S.~Ioffe and C.~Szegedy, ``Batch normalization: Accelerating deep network
  training by reducing internal covariate shift,'' in \emph{International
  conference on machine learning}.\hskip 1em plus 0.5em minus 0.4em\relax PMLR,
  2015, pp. 448--456.

\bibitem{Maas13rectifiernonlinearities}
A.~L. Maas, A.~Y. Hannun, and A.~Y. Ng, ``Rectifier nonlinearities improve
  neural network acoustic models,'' in \emph{in ICML Workshop on Deep Learning
  for Audio, Speech and Language Processing}, 2013.

\bibitem{scheibler2018pyroomacoustics}
R.~Scheibler, E.~Bezzam, and I.~Dokmani{\'c}, ``Pyroomacoustics: A python
  package for audio room simulation and array processing algorithms,'' in
  \emph{2018 IEEE International Conference on Acoustics, Speech and Signal
  Processing (ICASSP)}.\hskip 1em plus 0.5em minus 0.4em\relax IEEE, 2018, pp.
  351--355.

\bibitem{allen1979image}
J.~B. Allen and D.~A. Berkley, ``Image method for efficiently simulating
  small-room acoustics,'' \emph{The Journal of the Acoustical Society of
  America}, vol.~65, no.~4, pp. 943--950, 1979.

\bibitem{panayotov2015librispeech}
V.~Panayotov, G.~Chen, D.~Povey, and S.~Khudanpur, ``Librispeech: an asr corpus
  based on public domain audio books,'' in \emph{2015 IEEE international
  conference on acoustics, speech and signal processing (ICASSP)}.\hskip 1em
  plus 0.5em minus 0.4em\relax IEEE, 2015, pp. 5206--5210.

\bibitem{adam}
D.~P. Kingma and J.~Ba, ``Adam: A method for stochastic optimization,'' in
  \emph{International Conference on Learning Representations (ICLR)}, 2015.

\bibitem{omologo1997use}
M.~Omologo and P.~Svaizer, ``Use of the crosspower-spectrum phase in acoustic
  event location,'' \emph{IEEE Transactions on Speech and Audio Processing},
  vol.~5, no.~3, pp. 288--292, 1997.

\bibitem{wang2007robust}
L.~Wang, N.~Kitaoka, and S.~Nakagawa, ``Robust distant speaker recognition
  based on position-dependent cmn by combining speaker-specific gmm with
  speaker-adapted hmm,'' \emph{Speech communication}, vol.~49, no.~6, pp.
  501--513, 2007.

\bibitem{plinge2016acoustic}
A.~Plinge, F.~Jacob, R.~Haeb-Umbach, and G.~A. Fink, ``Acoustic microphone
  geometry calibration: An overview and experimental evaluation of
  state-of-the-art algorithms,'' \emph{IEEE Signal Processing Magazine},
  vol.~33, no.~4, pp. 14--29, 2016.

\bibitem{velasco2016tdoa}
J.~Velasco, D.~Pizarro, J.~Macias-Guarasa, and A.~Asaei, ``Tdoa matrices:
  Algebraic properties and their application to robust denoising with missing
  data,'' \emph{IEEE Transactions on signal processing}, vol.~64, no.~20, pp.
  5242--5254, 2016.

\end{thebibliography}

\end{document}